\documentclass[universe,article,accept,pdftex,moreauthors]{Definitions/mdpi} 

\firstpage{1} 
\pubvolume{1}
\issuenum{1}
\articlenumber{0}
\pubyear{2025}
\copyrightyear{2025}
\externaleditor{Firstname Lastname}
\datereceived{3 February 2025} 
\daterevised{20 February 2025} 
\dateaccepted{25 February 2025} 
\datepublished{ } 
\hreflink{https://doi.org/} 

\DeclareRobustCommand{\ion}[2]{%
\relax\ifmmode
\ifx\testbx\f@series
{\mathbf{#1\,\mathsc{#2}}}\else
{\mathrm{#1\,\mathsc{#2}}}\fi
\else\textup{#1\,{\mdseries\textsc{#2}}}%
\fi}

\Title{Solid Identification of Extragalactic Gamma-Ray Source Using High-Resolution Radio Interferometric Observation}

\TitleCitation{Solid Identification of Extragalactic Gamma-Ray Source Using High-Resolution Radio Interferometric Observation}

\Author{Krisztina 
 \'Eva Gab\'anyi  $^{1,2,3,4,5,}$*\orcidA{}, S\'andor Frey $^{3,4,6}$\orcidB{}, Krisztina Perger $^{3,4}$\orcidC{} and Emma Kun $^{3,4,7,8,9}$\orcidD{}} 

\AuthorNames{Krisztina \'Eva Gab\'anyi, S\'andor Frey, Krisztina Perger and Emma Kun}

\AuthorCitation{Gab\'anyi, K.\'E.; Frey, S.; Perger, K.; Kun, E.}

\address{%
$^{1}$ \quad Department of  Astronomy, Institute of Physics and Astronomy, ELTE E\"otv\"os Lor\'and University,
              P\'azm\'any P\'eter S\'et\'any 1/A, 1117 Budapest, Hungary\\
$^{2}$ \quad HUN-REN--ELTE Extragalactic Astrophysics Research Group, ELTE E\"otv\"os Lor\'and University,
              P\'azm\'any P\'eter S\'et\'any 1/A, 1117 Budapest, Hungary\\
$^{3}$ \quad Konkoly Observatory, HUN-REN Research Centre for Astronomy and Earth Sciences, Konkoly-Thege Mikl\'os \'ut 15-17, 1121 Budapest, Hungary; frey.sandor@csfk.org (S.F.); perger.krisztina@csfk.org (K.P.); kun.emma@csfk.org (E.K.) 
\\
$^{4}$ \quad CSFK, MTA Centre of Excellence, Konkoly-Thege Mikl\'os \'ut 15-17, 1121 Budapest, Hungary\\

$^{5}$ \quad Institute of Astronomy, Faculty of Physics, Astronomy and Informatics, Nicolaus Copernicus University, Grudziądzka 5, 87-100 Toru\'n, Poland\\
$^{6}$ \quad Institute of Physics and Astronomy, ELTE E\"otv\"os Lor\'and University,
              P\'azm\'any P\'eter S\'et\'any 1/A, 1117~Budapest, Hungary\\
$^{7}$ \quad Theoretical Physics IV: 
 Plasma-Astroparticle Physics, Faculty for Physics \& Astronomy, Ruhr University Bochum, 44780 Bochum, Germany\\
$^{8}$ \quad Ruhr Astroparticle and Plasma Physics Center, Ruhr-Universit\"at Bochum, 44780 Bochum, Germany\\
$^{9}$ \quad Astronomical Institute, Faculty for Physics \& Astronomy, Ruhr-Universität Bochum, 44780 Bochum, Germany}

\corres{Correspondence: k.gabanyi@astro.elte.hu}

\abstract{The dominant fraction of the extragalactic $\gamma$-ray sources are blazars, active galactic nuclei with jets inclined at
a small angle to the line of sight. Apart from blazars, a few dozen narrow-line Seyfert 1 galaxies (NLS1) and a number of radio galaxies are associated with $\gamma$-ray sources. The identification of $\gamma$-ray sources requires multiwavelength follow-up observations since several candidates could reside within the relatively large $\gamma$-ray localisation area. The $\gamma$-ray source 4FGL\,0959.6$+$4606 
 was originally associated with a radio galaxy. However, follow-up multiwavelength work 
suggested a nearby NLS1 as the more probable origin of the $\gamma$-ray emission. We performed high-resolution very long baseline interferometry (VLBI) observation at $5$\,GHz of both proposed counterparts of 4FGL\,0959.6$+$4606. We clearly detected the NLS1 source SDSS\,J095909.51$+$460014.3 
 with relativistically boosted jet emission. On the other hand, we did not detect milliarcsecond-scale compact emission in the radio galaxy 2MASX\,J09591976$+$4603515. Our VLBI imaging results suggest that the NLS1 object is the origin of the $\gamma$-ray emission in 4FGL\,0959.6$+$4606.}


\begin{document}



\section{Introduction}
\label{sec:intro}
Observations with a Large Area Telescope (LAT) of the \textit{Fermi} 
 satellite revealed that the extragalactic $\gamma$-ray sky is dominated by blazars, i.e.,~radio-loud active galactic nuclei (AGN) whose jets are directed at a small angle to the line of sight~\cite{Ajello2020}. According to the reclassification~\cite{Foschini2022} of the second data release of the fourth \textit{Fermi}--LAT catalog~\cite{Abdollahi2020}, $63\%$ of the $2980$ $\gamma$-ray-emitting AGN (outside of the Galactic Plane, with~galactic latitudes $|b|>10^\circ$) are blazars. The~remaining sources are associated with other AGN types, e.g.,~misaligned AGN (including radio galaxies) and narrow-line Seyfert 1 (NLS1) galaxies. NLS1 galaxies were the third AGN type, after BL Lac objects and flat-spectrum radio quasars (FSRQs), that were recognised to be able to produce $\gamma$-ray emission~\cite{Abdo2009}.
There are $24$ $\gamma$-ray-emitting NLS1 galaxies and a few ambiguous or intermediate cases reported in~\cite{Foschini2022}. Apart from those, additional $\gamma$-ray emitter sources are connected to NLS1 galaxies in the literature, such as 4C$+$04.42~\cite{Romano2018, Yao2015}, 4FGL\,J0117.9$+$1430~\cite{2024MNRAS_UGS}, 3FGL\,J0031.6$+$0938 
~\cite{Paiano2019}, and~J164100.10$+$345452.7~\cite{Romano2018, Lahteenmaki2018}.

NLS1 galaxies are characterised by narrow permitted emission lines (with a full width at half-maximum, FWHM(H$\beta$)$ <2000\,\textrm{km\,s}^{-1}$, \cite{Goodrich1989}), strong \ion{Fe}{ii} multiplets, and weak \ion{O}{[iii]} emission~\cite{Osterbrock1985}. Most of the NLS1s are radio-quiet, as only $\sim$7\% of them show radio emission~\cite{Komossa2006, Rakshit2021}. Radio studies accomplished at kpc-scale resolutions showed diverse radio morphologies of NLS1 sources and, combined with multiwavelength data, revealed that the radio emission may originate from the AGN and/or the star formation in the host galaxy~\cite{Jarvela2022}.

On the other hand, the~extremely radio-loud NLS1s ($\sim$2.5\%) show blazar-like characteristics: flat radio spectrum, compact cores on milliarcsecond (mas) scale, high brightness temperatures, and~significant variability, see, e.g., 
 \cite{Kozak}. In~the Monitoring of Jets in Active Galactic Nuclei with Very Long Baseline Array Experiments (MOJAVE) program, three of the five monitored $\gamma$-ray-loud NLS1 sources also showed apparent superluminal jet component motions~\cite{Lister2016}. It was shown that radio-loud NLS1s with blazar-like characteristics often experience enhanced variability in the infrared, a~significant fraction also on intra-day {{timescales}, indicating that at least part of the infrared emission in these objects originates from the boosted synchrotron jet~\cite{Jiang2012,Gabanyi2018,Mao2021}. Short-timescale variability  was also reported in the optical band (intra-night optical variability) and~was also predominantly measured in $\gamma$-ray-emitting, jetted NLS1s, implying the relativistically boosted jet as the origin of the optical variability (e.g., \cite{Kshama2017, Ojha2022}). Concerning the higher-energy regime, enhanced X-ray variability in relatively radio-fainter (mildly radio-loud) NLS1 sources, in~comparison to regular Seyfert galaxies, may also indicate that they harbour jets contributing to their radio emission~\cite{Boller1996, Chainakun2017, Lister2018}. On~the other hand, the~rapid X-ray variability of the $\gamma$-ray-emitting NLS1, 1H0323$+$342, has been attributed to its accretion disk~\cite{Yao2015_1H}.

NLS1 galaxies are thought to contain a low-mass ($<$10$^8$\,M$_\odot$) supermassive black hole in their central engine and/or are often considered to be at an earlier evolutionary stage~\cite{Mathur2000,Komossa2006}. In~that context, radio-bright NLS1s showing blazar-like characteristics can be the low-mass counterparts of FSRQs~\cite{Foschini2020}. An~alternative scenario explains the narrowness of the permitted spectral lines by a broad line region with special (flattened) geometry~\cite{Decarli2008}.

In the \textit{Fermi}--LAT Fourth Source Catalog (4FGL) \cite{Abdollahi2020}, the~$\gamma$-ray-emitting object 4FGL\,0959.6$+$4606 was associated with the radio galaxy 2MASX\,J09591976$+$4603515 (hereafter 2MASX\,J0959$+$4603) at a redshift of $z=0.148$ \cite{redshift_03}. However, another probable radio-emitting counterpart, SDSS\,J095909.51$+$460014.3 (hereafter SDSS\,J0959$+$4600), an~NLS1 galaxy at $z=0.39892$ \cite{Rakshit2017}, was reported later by~\cite{Li2023}. According to~\cite{Jiang2012}, the~radio-loudness factor of SDSS\,J0959$+$4600 ($1000$) is two orders of magnitude larger compared to that of 2MASX\,J0959$+$4603 ($20$). Both radio sources are within the $95$\% confidence level $\gamma$-ray localisation uncertainty region if all \textit{Fermi}--LAT data are considered. 
However, if~the $15$-month data are used when 4FGL\,0959.6$+$4606 shows a high $\gamma$-ray state, then only SDSS\,J0959$+$4600 falls into the localisation radius. Additionally, SDSS\,J0959$+$4600 showed $2.5$\,mag brightening in infrared bands coincident in time with the $\gamma$-ray brightening, and~its spectral energy distribution (SED) could be well described with a single-zone homogeneous leptonic jet model. Thus, its multi-band characteristics were similar to the other known $\gamma$-ray-emitting NLS1 galaxies~\cite{Li2023}.

Since $\gamma$-rays in AGN are known to be produced in relativistic plasma jets expelled from the vicinity of the central supermassive black hole, and~because the high-resolution technique of very long baseline interferometry (VLBI) is the only method suitable for directly imaging the synchrotron radio emission of milliarcsecond (mas)-scale compact jets, we initiated observation of both proposed counterparts of 4FGL\,0959.6$+$4606 with the European VLBI Network (EVN). This way, we were able look for possible jets, compact radio-emitting features in them, and~signatures of relativistic beaming effects. Section~\ref{sec:obs} below describes the observation and data analysis. Section~\ref{sec:res} presents the results. We discuss our findings in Section~\ref{sec:disc} and conclude the paper in Section~\ref{sec:conc}.

Assuming a $\Lambda$ Cold Dark Matter cosmological model with a Hubble constant of\linebreak $H_0=70$\,km\,s$^{-1}$\,Mpc$^{-1}$ and~density parameters of $\Omega_\textrm{M}=0.3$ and $\Omega_\Lambda=0.7$ (the approximate values of the ones derived from the European Space Agency Planck satellite mission data~\cite{Planck}), $1$\,mas angular size corresponds to $5.36$\,kpc projected linear size at the redshift of J0959$+$4600~\cite{cosmocalc}.


\section{Observation and Data~Reduction}
\label{sec:obs}

Our EVN observation was conducted on 28 May 2023 at $5$\,GHz frequency, with~a bandwidth of $256$\,MHz divided into eight intermediate frequency bands (IFs), each containing $64$ channels. The~following antennas participated and provided useful data: Effelsberg (Germany), Tianma (China), Jodrell Bank Mark 2 (United Kingdom), Westerbork (The Netherlands), Medicina (Italy), Onsala (Sweden), Toru\'n (Poland), Yebes (Spain), and~Irbene (Latvia). Westerbork only recorded data in the four IFs at the upper half of the~band.

The observation lasted for $3$~h and was carried out in phase-reference mode~\cite{p-ref}. In~that mode, the~visibility phases are corrected using the measurements of a bright and compact calibrator source located at small angular separation from the target, and~the obtained phase solutions are transferred to the target source(s). In~our experiment, the~phase-calibrator source was ICRF\,J095819.6 $+$ 472507, located at $1.42^\circ$ angular separation from the target field. Its coordinates according to the latest, third realisation of the International Celestial Reference Frame (ICRF3) \cite{icrf3} are right ascension $\alpha_\mathrm{cal}=
09{^\mathrm{h}} 58^{\mathrm{m}} 19.671647^{\mathrm{s}} \pm 0.05$\,mas and declination $\delta_\mathrm{cal}= 47^{\circ} 25^{\prime} 07.84241^{\prime\prime} \pm 0.03$\,mas (note: 
 data obtained from \url{https://hpiers.obspm.fr/icrs-pc/newwww/icrf/icrf3sx.txt}, accessed on 2 February 2025).     

The angular separation of the two proposed counterparts is $\sim$2.5$^{\prime}$, comparable to the half-power beam width of the largest-diameter antennas of the array, Effelsberg ($\sim$2.4$^{\prime}$) and Tianma ($\sim$3.7$^{\prime}$), at this observing frequency. For~all other antennas, both targets fall well within their primary beams. Therefore, we asked for multi-phase-centre correlation at the SFXC correlator of the Joint Institute for VLBI European Research Infrastructure Consortium~\cite{sfxc}. During~the experiment, all antennas except for Effelsberg and Tianma observed the midpoint between the two proposed counterparts, while Effelsberg and Tianma were performing nodding-style phase referencing for both individual target sources. In~summary, the~phase-calibrator source and the target field were observed alternately by spending $\sim$1.3 min on the calibrator and $\sim$3.5 min on the target field by most of the antennas, while Effelsberg and Tianma observed 2MASX\,J0959$+$4603 and SDSS\,J0959$+$4600 alternately during the target-field scans. That way, the~smaller antennas had longer on-source times on the targets than the two largest antennas. The on-source time for both sources was $\sim$1.9 h.

The VLBI visibility data were reduced following the standard steps in the Astronomical Image Processing System ({\sc aips}
, \cite{aips}). These included the a priori amplitude calibration, ionospheric, parallactic angle, and instrumental delay corrections. Afterwards, fringe-fitting was performed on the phase-reference source, and~four-four channels were removed for all sources from the beginning and the end of each IF, to~avoid the lower amplitude values at the edges of the IFs. Then, the calibrated data of the phase-reference source were exported and imaged in the {\sc Difmap} program~\cite{Difmap}. During~the hybrid mapping process, several iterations of {\sc clean} deconvolution~\cite{clean} and phase self-calibration were performed. Then, the antenna-based amplitude correction factors were determined. These were generally below $10\%$, except~for the second IF of Tianma. The~amplitude corrections exceeding $5$\% were applied to all visibility data in {\sc aips}, and~a second fringe-fit was performed for the phase-reference calibrator, taking into account its structure recovered in the hybrid imaging. These solutions were interpolated and applied for the two target sources. Then, their data were exported to {\sc Difmap} for imaging. SDSS\,J0959$+$4600 turned out to be bright enough that we were able to perform fringe-fitting {\sc aips} on this target source, as~well.

\section{Results}
\label{sec:res}
\unskip

\subsection{The NLS1 Source SDSS\,J0959$+$4600}

The NLS1 source (SDSS\,J0959$+$4600) was clearly detected as a compact bright feature with an extension to the northwest (Figure\,\ref{fig:NLS_FF}). The~phase-referenced ICRF3 coordinates of the brightness peak are right ascension $\alpha_{\mathrm{VLBI}}=09^\mathrm{h} 59^\mathrm{m} 09.51304^\mathrm{s} $ and declination $\delta_{\mathrm{VLBI}}=46^{\circ} 00^{\prime} 14.2755^{\prime\prime}$. The~uncertainties of these coordinates are estimated to be $\sim$0.2 mas, dominated by the effect arising from the angular distance between the target and the phase-reference source~\cite{Kirsten}. 

\begin{figure}[H]
\includegraphics[width=10.5 cm]{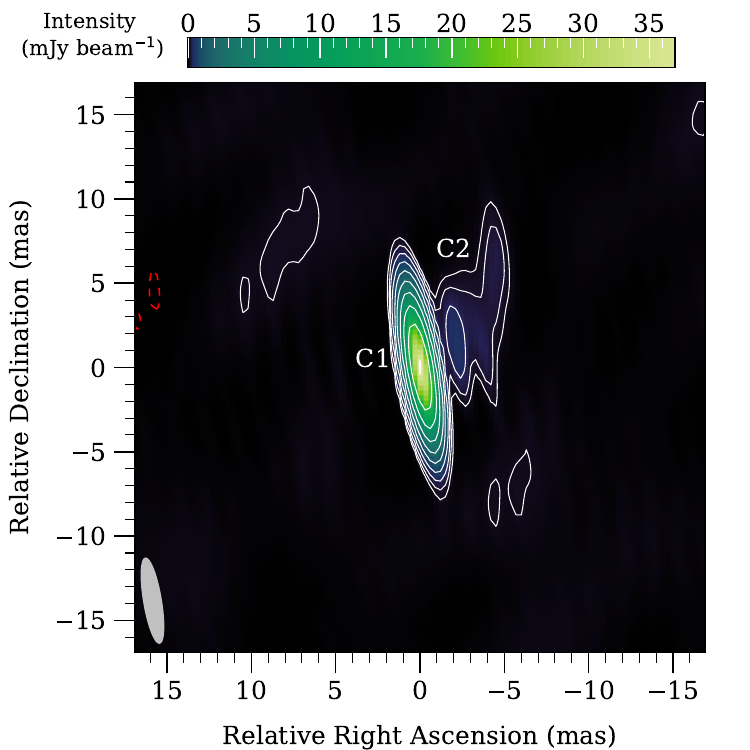}
\caption{Naturally weighted 
 $5$ GHz EVN map of SDSS\,J0959$+$4600 obtained after fringe-fitting to the target source data. The~two components are labeled. The~peak intensity is $36.9$\,mJy\,beam$^{-1}$. The~lowest contours are drawn at $\pm0.07$\,mJy\,beam$^{-1}$ (at $4\sigma$ image noise), and the~positive contour levels increase by a factor of $2$. The~elliptical Gaussian restoring beam size is $5.11$\,mas $\times~1.04$\,mas (FWHM) with a major axis position angle at $10^\circ$, and~it is shown in the lower left~corner. Red, dashed line indicates the negative contour. \label{fig:NLS_FF}}
\end{figure}

The maps obtained from the fringe-fitted and phase-referenced data were consistent regarding the source structure. The~peak intensity of the map created from the phase-referenced data was only marginally smaller, $36.1$\,mJy\,beam$^{-1}$, than that of the fringe-fitted dataset, $36.9$\,mJy\,beam$^{-1}$, while the $1 \sigma$ image noises of both maps were the same. The~total flux densities in the clean components, $\sim$40 mJy, agreed, as well. In~the subsequent part, we only discuss the fringe-fitted~data. 

To quantitatively describe the brightness distribution, we fitted the self-calibrated visibilities with circular Gaussian model components~\cite{modelfit}. Two components, C1 and C2, were needed to adequately describe the data. Their parameters, flux densities, FWHM sizes, and positional offsets from the centre of the map in the right ascension and declination directions are listed in Table~\ref{tab:model}. Flux density errors were calculated according to~\cite{Schinzel_phd},~including an additional $5\%$ flux density error in quadrature, to~account for the VLBI amplitude calibration uncertainty. The~errors of the offset position are estimated as $20\%$ of the restoring beam size in the corresponding direction~\cite{Lister2019}. The~size errors were estimated using modelfits to slightly differently self-calibrated datasets. The~size of C1 is close to but larger than the smallest resolvable size, $\sim$0.1 mas (as defined in~\cite{Kovalev2005}).

The separation of the two components is $\sim$4.3 mas, corresponding to a projected linear size of $\sim$23 pc.

\begin{table}[H] 
\caption{The parameters of the circular Gaussian model components fitted to the visibilities of SDSS\,J0959$+$4600. In~the last column, the~offsets from the centre of the map (from the bright C1 component) are given in the right ascension and declination directions, respectively. \label{tab:model}}
\begin{tabularx}{\textwidth}{cCCCC}
\toprule
\textbf{ID}	& \textbf{Peak Intensity (mJy\,beam$^{-1}$)}	& \textbf{Flux Density (mJy)}	& \textbf{FWHM Size (mas)} & \textbf{Offset (mas)}\\
\midrule
C1		& $36.79 \pm 0.06$ & $37.2 \pm 2.3$			& $0.14\pm 0.04 $ & -- \\
C2		& $0.73 \pm 0.03$ & $1.0 \pm 0.3$			& $0.7\pm 0.1$ & $-2.9 \pm 0.2,\linebreak 3.2\pm 1.0$ \\
\bottomrule
\end{tabularx}
\end{table}
\unskip

\subsection{The {Radio Galaxy 2MASX\,J0959$+$4603}}

At the position of the radio galaxy 2MASX\,J0959$+$4603, the~other proposed counterpart of the \textit{Fermi} $\gamma$-ray source 4FGL\,0959.6$+$4606, we did not detect compact radio emission in our EVN observation down to the rms noise level of $0.02$\,mJy\,beam$^{-1}$ ($1\sigma$). 
When we excluded the longest baselines (to Tianma) of the array, an~emission feature could be tentatively detected at the $\sim$5$\sigma$ level, with~a $\sim$0.1 mJy\,beam$^{-1}$ brightest pixel at $202$\,mas and $585$\,mas offset in right ascension and declination, respectively, from~the position of the radio galaxy as given in the catalog of the Wide-field Infrared Survey Explorer ({\it WISE}, \cite{wise}), AllWISE (\cite{2014yCat.2328....0C}, \url{https://cdsarc.cds.unistra.fr/viz-bin/cat/II/328}, accessed on 1 February 2025). According to~\cite{wise}, the~positional accuracy of the {\it WISE} is $\sim$150 mas.

\section{Discussion}
\label{sec:disc}

The most accurate optical position of the NLS1 galaxy is provided in the latest, third Data Release (DR3) \cite{Gaia-DR3} of the {\it Gaia} space telescope~\cite{Gaia}, $\alpha_\mathrm{opt}=
09{^\mathrm{h}} 59^{\mathrm{m}} 09.513055^{\mathrm{s}} \pm 0.27$\,mas and $\delta_\mathrm{opt}= 46^{\circ} 00^{\prime} 14.27589^{\prime\prime} \pm 0.24$\,mas. The VLBI and \textit{Gaia} right ascensions agree within the errors. Formally, the~optical declination is slightly to the north compared to the VLBI position. However, their $0.4$\,mas positional difference is comparable to the combined uncertainty of the optical and VLBI declinations, $0.31$\,mas. Therefore, we conclude that there is no significant difference between the \textit{Gaia} optical and the radio interferometric positions of the brightest component of SDSS\,J0959$+$4600. Thus, we can safely assume that component C1 is the radio core of the NLS1 galaxy, and~C2 is a jet component. Similar pc-scale core--jet structures have been seen in other $\gamma$-ray-emitting NLS1s (e.g., \cite{Kozak, Lister2016, Lister2018} and references therein).

The brightness temperature of a Gaussian component can be calculated as, e.g., \cite{tb_veres}:
\begin{equation}
    T_\mathrm{B}=1.22 \times10^{12} \frac{S}{\nu^2\theta^2} (1+z) \mathrm{\,K},
\end{equation}
where $S$ is the flux density in Jy, $\nu$ the observing frequency in GHz, $\theta$ the component size (FWHM) in mas, and~$z$ the redshift. In~the case of SDSS\,J0959$+$4600, \mbox{$T_\mathrm{B}^\mathrm{C1}=(12.6\pm3.7)\times 10^{10}$\,K}. This value exceeds the equipartition brightness temperature limit, $\approx 5\times10^{10}$\,K~\cite{equipartition}, indicating mild relativistic boosting in the source with a Doppler factor of $\delta=2.5 \pm 0.7$. This Doppler factor is much lower than the one assumed by~\cite{Li2023} ($\delta=13.5$) to fit the SED of SDSS\,J0959$+$4600. However, they modelled the high $\gamma$-ray flux state of the source, which occurred around September 2017, while our EVN observation took place $6$\,years~later.

While 2MASX\,J0959$+$4603 was not detected in our $5$-GHz VLBI observation, both objects have detections at GHz frequencies at lower resolutions: at $1.4$\,GHz in the Faint Images of the Radio Sky at Twenty centimeters survey (FIRST,~\cite{first}) and at $3$\,GHz in the more recent Very Large Array Sky Survey (VLASS,~\cite{vlass}). They appear as unresolved sources in both surveys; their flux densities are given in Table~\ref{tab:lowres}. The~flux density values of the first two VLASS epochs were obtained from the Quick Look catalogs 
 (version 3 for the first epoch and version 2 for the second epoch) downloaded from the Canadian Astronomy Data Centre (CADC) (note: 
 { \url{https://www.cadc-ccda.hia-iha.nrc-cnrc.gc.ca/en/vlass/}, accessed on 2 February 2025). For~SDSS\,J0959$+$4600, two objects were listed in both epochs. Following the CIRADA: VLASS Catalog User Guide (note: {\url {https://ws.cadc-ccda.hia-iha.nrc-cnrc.gc.ca/files/vault/cirada/tutorials/CIRADA__VLASS_catalogue_documentation_2023_june.pdf}, accessed on 2 February 2025), we used the one with duplicate flag value $1$. For~the third VLASS epoch, for~which no catalog has been produced yet, we downloaded the cutouts of the quick-look images from the CADC and fitted them using the {\sc aips} task {\sc imfit}. The~integrated flux density values obtained by the fitting were increased by $3\%$ since, according to the CIRADA: VLASS Catalog User Guide, the~flux densities are underestimated by this amount. Since SDSS\,J0959$+$4600 appeared on two cutouts, we took the average of the two fitted flux~densities.

\begin{table}[H] 
\caption{Radio flux densities of the two targets measured in low-resolution sky surveys with the VLA, from~the FIRST catalog~\cite{first}, and~as derived from the quick-look images of the VLASS~\cite{vlass}. See the text for~details. \label{tab:lowres}}
\begin{tabularx}{\textwidth}{CCCCC}
\toprule
\textbf{Name}	& \textbf{FIRST---1.4~GHz (mJy)}	& \multicolumn{3}{c}{\textbf{VLASS---3 GHz 
  (mJy)}} \\
 & 1997 Mar 21 
 & 2019 May 31 & 2021 Dec 5 & 2024  Jun~11 \\
\midrule
SDSS\,J0959$+$4600 
	& $27.1 \pm 1.4$			& $44.4\pm 1.1 $ & $32.4\pm0.2$ & $32.9\pm0.4$\\
2MASX\,J0959$+$4603		& $4.1 \pm 0.2$			& $2.2\pm 0.3$ & $2.2 \pm 0.2$ & $2.4\pm 0.3$ \\
\bottomrule
\end{tabularx}
\end{table}

Apart from the last VLASS epoch, flux density values were reported by~\cite{Li2023}. However, they used different software to fit the images, and~they may not have had access to the reprocessed first-epoch data at that time. With~the exception of the first-epoch VLASS flux density of SDSS\,J0959$+$4600, we obtained similar values for both sources; thus, we can confirm that 2MASX\,J0959$+$4603 did not show variability at $3$\,GHz. On~the other hand, the~flux density of SDSS\,J0959$+$4600 decreased significantly between the first two VLASS~epochs. 

We calculated the power-law spectral index $\alpha$ (defined as $S\propto\nu^\alpha$, where $\nu$ is the frequency) for 2MASX\,J0959$+$4603 between $1.4$\,GHz and $3$\,GHz using the average of the VLASS flux densities. The~obtained spectrum is steep, with $\alpha_{4603}=-0.8 \pm 0.3$. {\ Using this for extrapolation}, we would expect a total flux density of $\sim$1.6 mJy at $5$\,GHz for 2MASX\,J0959$+$4603. This should be easily detected in our EVN experiment if it is contained in a mas-scale compact feature. In~turn, its non-detection implies that the size of the radio-emitting region in 2MASX\,J0959$+$4603 exceeds the largest recoverable size of our observation, $\sim$50 mas. {The origin of this radio emission is not clear. It can be created in the extended lobe of the AGN or~arise from star formation in the host galaxy, or~it can be the result of the combination of both processes. If~we assume that the $1.4$-GHz flux density of 2MASX\,J0959$+$4603 solely arises from star formation, its radio power, $\sim$2.4~$\times10^{23}$\,W\,Hz$^{-1}$, would imply a star formation rate of $\sim$130 M$_\odot$\,yr$^{-1}$ \cite{Hopkins}. However, its infrared colors measured by the {\it WISE} satellite put it into the AGN locus of the infrared color--color diagram instead of the starburst region~\cite{Jarrett2011}.}

{For SDSS\,J0959$+$4600, the~FIRST and first-epoch VLASS flux density values imply a rather inverted spectrum. However, due to the apparent flux density variability, the~specific spectral index value may be misleading in that case, because~the observing epochs are different. Nevertheless, accepting it at face value, an~inverted radio spectrum would be consistent with a blazar-like, beamed NLS1 object.}
Assuming that we did not lose significant flux density in our VLBI measurement due to an extended structure resolved out at the long baselines, and~disregarding possible source variability, we could also calculate the spectral index of SDSS\,J0959$+$4600 between $3$\,GHz and $5$\,GHz. For~that, we used the last VLASS epoch, the~one closest to our EVN observation. The~obtained $\alpha=0.3 \pm 0.1$ implies that the radio spectrum is flat between these~frequencies.

\section{Conclusions}
\label{sec:conc}

The $\gamma$-ray source 4FGL\,0959.6$+$4606 was originally associated with a radio galaxy~\cite{Abdollahi2020}. However, ref.~\cite{Li2023} showed that an extremely radio-loud NLS1 galaxy is a more probable counterpart of the high-energy source. To~help resolve this issue using high-resolution radio imaging, we performed $5$ GHz VLBI observation of both sources with the EVN in 2023. Among~the two candidate counterparts of the $\gamma$-ray source, the~NLS1 SDSS\,J0959$+$4600 was clearly detected with a bright mas-scale compact core and a fainter jet feature. The~brightness temperature of the core exceeds the equipartition limit, indicating moderate Doppler boosting. On~the other hand, we did not convincingly detect any mas-scale compact radio feature in the radio galaxy 2MASX\,J0959$+$4603, indicating that its total radio emission originates from spatially extended features and not the core of an AGN. Our results lend strong support to the suggestion of~\cite{Li2023} that the radio-bright NLS1 object is the most probable source of the $\gamma$-ray emission in 4FGL\,0959.6$+$4606.

{ Our finding highlights the significance of high-resolution radio observations in the identification of $\gamma$-ray-emitting objects (see also~\cite{4C_2018}). It can be quite important when the different proposed counterparts belong to source groups represented in relatively small numbers among the known $\gamma$-ray emitters. Thus, like in the case of 4FGL\,0959.6$+$460, the~proposed counterparts, a~radio galaxy and an NLS1 object, both can have essential but~different implications for the possible $\gamma$-ray emission mechanisms. Our findings and the growing number of other known $\gamma$-ray-emitting NLS1 objects confirm that contrary to previous beliefs, these unique objects are not only capable of but~are proficient in launching powerful jets similar to blazars.}


\vspace{6pt} 




\authorcontributions{Conceptualisation, K.\'E.G., S.F., K.P., and~E.K.; formal analysis, K.\'E.G.; wri\mbox{ting---origi}nal draft preparation, K.\'E.G.; writing---review and editing: S.F., K.P., and~E.K.; visualisation, K.\'E.G. and K.P. All authors have read and agreed to the published version of the~manuscript.}

\funding{This 
 research was funded by the Hungarian National Research, Development and Innovation Office (NKFIH), grant number OTKA K134213, and~by the NKFIH excellence grant TKP2021-NKTA-64. This project has received funding from the HUN-REN Hungarian Research Network.}

\dataavailability{The calibrated VLBI data are available from the corresponding author upon reasonable request. The~raw VLBI data are available from the EVN Data Archive (\url{http://archive.jive.nl/scripts/portal.php}, accessed on 2 February 2025) under project code EG125.} 




\acknowledgments{{We thank the anonymous referees whose suggestions helped improve the~paper.} The EVN is a joint facility of independent European, African, Asian, and~North American radio astronomy institutes. Scientific results from data presented in this publication are derived from the following EVN project code: EG125. The~National Radio Astronomy Observatory is a facility of the National Science Foundation operated under cooperative agreement by Associated Universities, Inc. The~Canadian Initiative for Radio Astronomy Data Analysis (CIRADA) program is funded by a grant from the Canada Foundation for Innovation 2017 Innovation Fund (Project 35999), as~well as by the Provinces of Ontario, British Columbia, Alberta, Manitoba, and~Quebec. 
}

\conflictsofinterest{The authors declare no conflicts of interest. The~funders had no role in the design of the study; in the collection, analyses, or~interpretation of data; in the writing of the manuscript; or in the decision to publish the~results.} 



\abbreviations{Abbreviations}{
The following abbreviations are used in this manuscript:\\

\noindent 
\begin{tabular}{@{}ll}
AGN & active galactic nuclei\\
AIPS & Astronomical Image Processing System\\
CADC & Canadian Astronomy Data Centre\\
CIRADA & Canadian Initiative for Radio Astronomy Data Analysis\\
EVN & European VLBI Network\\
FIRST & Faint Images of the Radio Sky at Twenty centimeters\\
FSRQ & flat-spectrum radio quasar\\
FWHM & full width at half-maximum\\
ICRF3 & third realisation of the International Celestial Reference Frame\\
LAT & Large Area Telescope\\
MOJAVE & \textls[-15]{Monitoring of Jets in Active Galactic Nuclei with Very Long Baseline Array Experiments}\\
mas & milliarcsecond\\
NLS1 & narrow-line Seyfert 1\\
SED & spectral energy distribution\\
VLA & Karl G. Jansky Very Large Array\\
VLASS & Very Large Array Sky Survey\\
VLBI & very long baseline interferometry\\
WISE & Wide-field Infrared Survey Explorer\\
\end{tabular}
}



\begin{adjustwidth}{-\extralength}{0cm}

\reftitle{References}

\PublishersNote{}
\end{adjustwidth}
\end{document}